\documentclass[a4paper,twocolumn,11pt]{quantumarticle}
\pdfoutput=1
\usepackage[utf8]{inputenc}
\usepackage[english]{babel}
\usepackage[T1]{fontenc}
\usepackage{amsmath}
\usepackage{tikz}
\usepackage{lipsum}
\usepackage{xcolor}
\usepackage{soul}
\usepackage{physics,array}
\usepackage{multirow}

\usepackage[numbers]{natbib}
\usepackage{hyperref} 

\begin{document}

\title{Q-Pandora Unboxed: Characterizing Noise Resilience of Quantum Error Correction Codes}

\author{Avimita Chatterjee*}
\email{amc8313@psu.edu}
\affiliation{School of EECS, The Pennsylvania State University, USA}
\orcid{0009-0001-7421-9334}

\author{Subrata Das*}
\email{sjd6366@psu.edu}
\affiliation{School of EECS, The Pennsylvania State University, USA}

\author{Swaroop Ghosh}
\email{szg212@psu.edu}
\affiliation{School of EECS, The Pennsylvania State University, USA}

\let\oldthefootnote=\thefootnote
\renewcommand{\thefootnote}{\fnsymbol{footnote}}
\footnotetext[1]{Both authors contributed equally to this research.}
\let\thefootnote=\oldthefootnote

\maketitle

\begin{abstract}
  Quantum error correction codes (QECCs) are critical for realizing reliable quantum computing by protecting fragile quantum states against noise and errors. However, limited research has analyzed the noise resilience of QECCs to help select optimal codes. This paper conducts a comprehensive study analyzing two QECCs - rotated and unrotated surface codes - under different error types and noise models using simulations. Among them, rotated surface codes perform best with higher thresholds attributed to simplicity and lower qubit overhead. The noise threshold, or the point at which QECCs become ineffective, surpasses the error rate found in contemporary quantum processors. When confronting quantum hardware where a specific error or noise model is dominant, a discernible hierarchy emerges for surface code implementation in terms of resource demand. This ordering is consistently observed across unrotated, and rotated surface codes. Our noise model analysis ranks the code-capacity model as the most pessimistic and circuit-level model as the most realistic. The study maps error thresholds, revealing surface code's advantage over modern quantum processors. It also shows higher code distances and rounds consistently improve performance. However, excessive distances needlessly increase qubit overhead. By matching target logical error rates and feasible number of qubits to optimal surface code parameters, our study demonstrates the necessity of tailoring these codes to balance reliability and qubit resources. Conclusively, we underscore the significance of addressing the notable challenges associated with surface code overheads and qubit improvements. 
\end{abstract}


\section{Introduction}

Quantum computing uses quantum mechanics to perform tasks beyond classical computing's scope with applications in simulating molecules for drug discovery, enhancing finance optimization, accelerating machine learning, improving optimization tasks, and transforming supply chain management, to name a few \cite{reiher2017elucidating, orus2019quantum, gachnang2022quantum, schuld2015introduction, ajagekar2019quantum}. These developments open the door to revolutionary commercial applications. Quantum computing, though advancing, faces challenges like qubit stability and quantum noise \cite{mouloudakis2021entanglement, clerk2010introduction}. Error correction, critical due to inevitable qubit noise, is being innovated via Quantum Error Correction Codes (QECCs) \cite{devitt2013quantum} to realize fault-tolerant quantum computing \cite{preskill1998fault}. Classical error correction techniques \cite{hamming1950error} face hurdles in quantum computing due to the no-cloning theorem \cite{wootters2009no}, and wavefunction collapse during qubit measurement \cite{mackey2013mathematical}. 
Studies showcase quantum codes, such as the five-qubit code, Bacon–Shor code, topological code, surface code, color code, and heavy-hexagon code \cite{sundaresan2022matching, abobeih2022fault, bacon2006operator, kitaev1997quantum, krinner2022realizing, bombin2006topological}, capable of rectifying single errors. Despite advances, one key question remains: Can enlarging the error-correcting code reduce logical error rates in real devices? Theory suggests that higher code distances should lead to fewer logical errors. However, practical proof necessitates device scaling. QEC could significantly cut quantum processor error rates, but at the cost of time and qubit overhead \cite{knill1998resilient}.

\subsection{Motivation}

``The large-scale quantum machine, though it may be the computer scientist’s dream, is the experimenter’s nightmare'' (quoted from \cite{haroche1996quantum}). This challenge arises due to the inherent fragility of quantum states, which are extremely sensitive to any form of external noise or environmental interaction. Quantum information processing is prone to errors and noise obstructing consistent and reliable computation. The state-of-the-art quantum processors report around $10^{-3}$ error rates \cite{ballance2016high, huang2019fidelity, rol2019fast, jurcevic2021demonstration}. 
QECCs serve as a protective barrier against such errors, yet the efficacy of these codes is largely dictated by the nature and intensity of the encountered noise \cite{terhal2015quantum}. Thus the detailed characterization of QECCs under noise allows for a comprehensive understanding of how these codes function under varying noise conditions. This, in turn, is integral to the selection of the most suitable code for a particular quantum computing task. It not only leads to improved efficiency but also bolsters the reliability of quantum computations. 

\subsection{Contributions}

To the best of our knowledge, this paper represents the first in-depth analysis of QECCs under noise. Despite efforts made in earlier literature to explore the efficiency and scalability of QECCs, they often fall short of providing a comprehensive analysis with respect to noise \cite{tomita2014low, darmawan2017tensor, tuckett2020fault, tuckett2019tailoring, google2023suppressing}. In this research, we use STIM \cite{gidney2021stim}, a fast simulator for quantum stabilizer circuits. According to a recent study \cite{krinner2022realizing}, STIM does an exceptional job of mimicking the performance of QECCs on real quantum machines. As today's quantum computers do not have enough qubits to run QECCs directly, we find STIM to be the best tool for our study.

In most cases, however, a code capable of managing both bit-flip and phase-flip errors, like surface codes, is necessary. Among surface codes, our analysis favors rotated surface codes for their superior thresholds, attributed to less complexity and fewer qubit needs. In the context of quantum hardware with predominant error types or noise models, we can arrange the resource requirements of different codes from least to most demanding as follows: reset \& readout errors, depolarizing errors, and gate errors for error types; and the phenomenological model, the code-capacity model, and the circuit-level model for noise models. Our experiments reveal surface code's error threshold or the point where these codes stop working is at least ten times higher than existing quantum processors. They also have the capability of reducing the logical error rate significantly. This performance can be further enhanced by increasing resources like code distance and rounds, which we validate, to reduce the logical error rate. We also highlight that the highest-performing surface code is not always necessary. By understanding a system's targeted error rate and noise types, we can optimize surface code parameters to reduce surplus qubit and gate overhead. For an in-depth understanding and characterization of these QECCs, refer to the comprehensive analysis presented in Section \ref{advancing_eff}. We assume that readers have a foundational understanding of QEC including rudimentary principles of classical error correction, repetition codes, stabilizer \& encoding circuits, and topological codes, along with an understanding of the distinction between toric and surface codes. For such background information, one can refer to recent literature on the topic such as \cite{chatterjee2023quantum}.

\subsection{Structuring the Paper}

Overview of QECCs and literature review is provided in Sections \ref{qecc} and \ref{related_works}, respectively. 
In Section \ref{noise}, we examine noise in quantum computing and explain our experimental setup and the methodology. 
Section \ref{all_codes} offers a detailed analysis of surface codes in relation to the aforementioned types of noise and noise models. 
We delve into strategies for optimal and scalable code selection in Section \ref{advancing_eff}. 
Section \ref{limitaions} delineates the limitations and avenues for future exploration. 
Finally, Section \ref{conclusion} summarizes our findings.
\section{Overview of Quantum Error Correction} \label{qecc}

\subsection{Fundamentals}

Quantum Error Correction (QEC) employs Quantum Error Correction Codes (QECCs) to safeguard quantum states from errors during computations or transmissions. QECCs disperse quantum information across qubits and work in three phases: encoding, detecting errors, and correction. This section delves into rotated and unrotated surface codes, their structures, and error management methods. We will also discuss the significance of error rates and the concept of a threshold in QEC. Some core principles of QECCs are:

\begin{enumerate}
    \item First, Pauli operators (X, Y, Z) and their anti-commuting properties are fundamental for error identification and rectification. X and Z are anti-commuting, implying that $(X \otimes I)( I \otimes Z) = - (I \otimes Z) (X \otimes I)$. Consequently, error detection and correction hinge on this relationship, as X-type stabilizers are used to identify Z-type errors, and vice versa. 
    \item Second, ancilla (or helper) qubits help in detecting errors without measuring and potentially collapsing the quantum states of data qubits.
    \item Third, syndromes, derived from ancilla qubit measurements, help pinpoint errors that specify the nature of the error and the affected qubits.
    \item Fourth, A round in QECC represents a full cycle of error detection and correction. Maintaining a balance between the number of rounds and the code distance is essential.
    \item Finally, QECCs distinguish between `logical' and `physical' qubits. A physical qubit is the original state that is being encoded into multiple logical qubits to form a logical state. In general, QECCs are represented as $[[n,k,\delta]]$, where $n$ is the number of logical qubits in the logical state post encoding; $k$ is the original number of qubits, and $\delta$ is the distance of the code. 
\end{enumerate}

\subsection{Surface Codes}

The surface code \cite{dennis2002topological} is a two-dimensional QECC and is currently one of the most popular codes due to its high threshold for errors. It succeeded in the proposal of toric codes, marking it as one of the earliest instances of topological codes. The operation of the surface code involves encoding physical qubits within a logical space that spans two dimensions. This unique layout plays a crucial role in its ability to detect and correct errors efficiently, thus contributing to its high error threshold. A considerable number of experiments have been conducted to realize surface codes in practical applications \cite{krinner2022realizing, takita2017experimental}. The distinguishing attribute of the surface code emerges from its unique approach to error detection and correction. Ancilla qubits, crucial to this process, are employed to conduct measurements that identify both bit-flip and phase-flip errors: the two primary categories of quantum errors. This methodology aligns with the strategies used by most QECCs, where measurements are performed in the form of X and Z stabilizer checks. 

An n-distance surface code is expressed in the form of an $n \times n$ lattice, where each blob signifies a qubit integral to the logical state. The lattice is structured in a manner that ensures every qubit is subject to both X and Z stabilizers, tasked with detecting phase and bit-flip errors, respectively. Each stabilizer check is a product of several Pauli operators (X or Z) acting on a subset of the qubits. If an error occurs, it will change the outcome of the stabilizer checks associated with that qubit. By performing these checks, it is possible to identify when and where an error has occurred after using a decoding algorithm \cite{kolmogorov2009blossom, higgott2022pymatching}. After errors have been detected, they can be corrected using quantum gates to flip the affected qubits back to their correct states. 

Unrotated \cite{fowler2012surface} and rotated \cite{tomita2014low} surface codes are two different layouts for implementing surface code QEC. The difference lies primarily in their lattice structures and the arrangement of qubits, which influence their error-correction capabilities.

\begin{figure}
    \centering
    \includegraphics[width=0.7\linewidth]{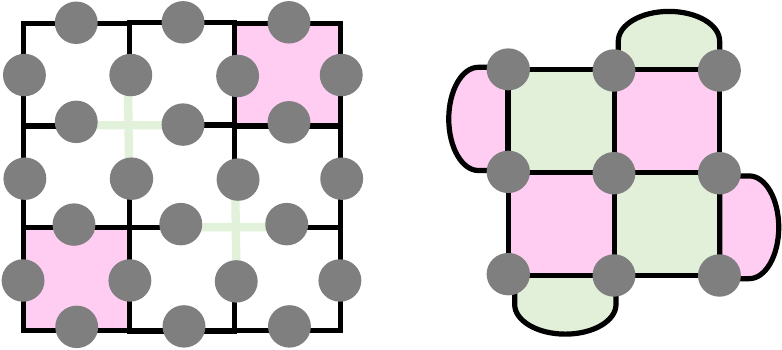}
    \caption{\textbf{Representation of distance $3$ surface codes.} Unrotated surface code \textit{(left)} and rotated surface code \textit{(right)}. The grey-blob depicted qubits are acted upon by pink-surface represented Z-stabilizers and green-surface indicated X-stabilizers.
    }
    \label{fig:both_surface_codes}
    \vspace{-10pt}
\end{figure}

\begin{enumerate}
    \item \textit{Unrotated surface codes:} These utilize a square lattice arrangement of qubits. Here, the data qubits form the edges of squares with each surface or plaquette corresponding to a Z-stabilizer influencing all four qubits of the edges. X-stabilizers, represented by the vertices, act on the qubits that lie on the edges connected to the vertex. Fig. \ref{fig:both_surface_codes} \textit{(left)} depicts an unrotated surface code, with grey-blob denoted qubits acted upon by pink-square represented Z-stabilizers and green-line depicted X-stabilizers. 
    \item \textit{Rotated Surface Codes:} These deploy a tilted lattice pattern where X-stabilizers and Z-stabilizers occupy alternating boxes of a lattice, akin to a chessboard layout. These stabilizers operate on the qubits positioned on the vertices of the specific box. Fig. \ref{fig:both_surface_codes} \textit{(right)} exemplifies distances 3 rotated surface codes. Here, pink-surfaces symbolize Z-stabilizers, while green-surfaces denote X-stabilizers. Collectively, these stabilizers act on all the grey-blob illustrated qubits that constitute the logical state.
\end{enumerate}

Generally, rotated surface codes are favored for achieving robust long-distance QEC due to their slightly higher threshold and less complex implementation. Nevertheless, both these configurations necessitate a substantial number of physical qubits and sophisticated control, making the realization of a comprehensive, fault-tolerant quantum computer using surface codes a formidable challenge for quantum engineering.

\begin{figure}
    \centering
    \includegraphics[width=1\linewidth]{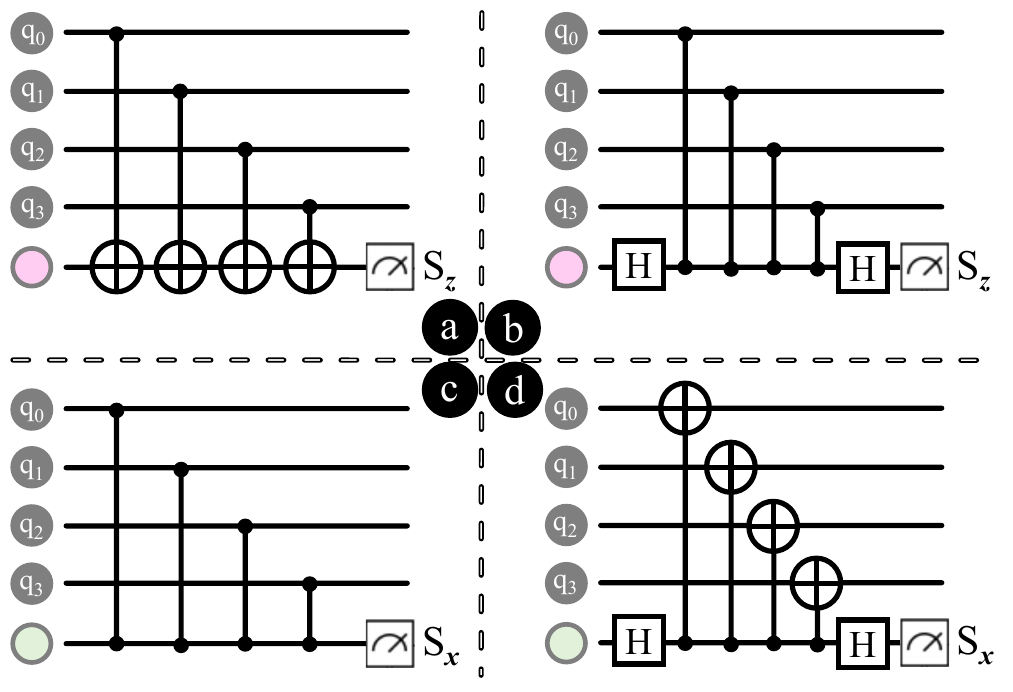}
    \caption{\textbf{Schematic representation of Z and X stabilizer construction.} \raisebox{.5pt}{\textcircled{\raisebox{-.4pt} {\textbf{a}}}} and \raisebox{.5pt}{\textcircled{\raisebox{-.9pt} {\textbf{b}}}} display the formation of Z-stabilizers using CNOT and CZ with Hadamard gates respectively, while \raisebox{.5pt}{\textcircled{\raisebox{-.2pt} {\textbf{c}}}} and \raisebox{.5pt}{\textcircled{\raisebox{-.9pt} {\textbf{d}}}} depict X-stabilizers formed using CZ and CNOT with Hadamard gates respectively. 
    }
    \label{fig:all_stabs}
    \vspace{-10pt}
\end{figure}

When a stabilizer operates on a set of four qubits ($q_0, q_1, q_2, q_3$), there are typically two methods to represent this interaction in the form of a quantum circuit. Figure \ref{fig:all_stabs} demonstrates two distinct approaches to constructing X and Z stabilizers, either with or without Hadamard gates. \raisebox{.5pt}{\textcircled{\raisebox{-.4pt} {\textbf{a}}}} Depicts the formation of a Z-stabilizer utilizing solely CNOT gates. The outcome of the stabilizer is projected onto an ancillary qubit, signified by a pink-blob, and subsequently measured to provide the syndrome measurement $S_z$. \raisebox{.5pt}{\textcircled{\raisebox{-.9pt} {\textbf{b}}}} Exhibits the same Z-stabilizer circuit, but this time constructed using CZ and Hadamard gates. The stabilizer's output is projected onto the pink-blob denoted ancillary qubit and is later measured as $S_z$. \raisebox{.5pt}{\textcircled{\raisebox{-.2pt} {\textbf{c}}}} Showcases an X-stabilizer constructed using CZ gates. It projects its value onto an ancillary qubit denoted as a green-blob, which is then measured as the syndrome value of $S_x$. \raisebox{.5pt}{\textcircled{\raisebox{-.9pt} {\textbf{d}}}} Presents the same X-stabilizer but designed using CNOT and Hadamard gates. The stabilizer's output is projected onto the green-blob represented ancillary qubit and subsequently measured as $S_x$. In all cases, the projection of the stabilizer's value onto the ancillary qubits is critical for error detection without disturbing the logical state.

\vspace{-20pt}
\subsection{Threshold of Quantum Error Correction Codes} 

The physical error rate measures how often errors arise from quantum operations like gates, measurements, and qubit initialization. A high rate implies many faulty operations, compromising quantum computation results. Conversely, the logical error rate evaluates errors in logical qubits, derived from multiple physical qubits through QECCs. Ideally, logical rates should be lower, as QECCs detect and rectify physical errors. However, an excessively high physical error rate or a limited QECC might still allow errors in logical qubits.

A primary objective of QECCs is minimizing the logical error rate, even amidst high physical error rates. The `threshold' error rate, a crucial performance metric, denotes the maximum physical error rate a QECC can handle while still diminishing the logical rate. Beyond this threshold, the code's redundancy can't adequately counteract all errors due to the sheer volume. Hence, surpassing the threshold necessitates an alternative error correction approach or enhanced quantum systems. Performance evaluations typically center on rates beneath this threshold to guide optimal QECC design.
\section{Related Works} \label{related_works}


In a seminal work by Tomita and Svore, they investigated the experimental implementation of a distance-3 surface code under realistic quantum noise conditions \cite{tomita2014low}. Their simulations, which were conducted under amplitude and phase damping, suggested that the Pauli-twirl approximation tends to give a pessimistic threshold estimate. While their research provided valuable insights into this specific type of surface code, it lacked a broader exploration of the noise resilience of various QECCs.

Darmawan and Poulin took a methodologically distinct approach by proposing a tensor network algorithm for simulating the surface code under arbitrary local noise \cite{darmawan2017tensor}. Their work stood out because it addressed the challenges of simulating the surface code under realistic conditions, whereas many previous studies had relied on overly simplified noise models. Despite the novelty of their method, their research was exclusively centered on the surface code, not extending its scope to other QECCs.

Tuckett et al. made significant contributions by addressing biased noise challenges in quantum computing \cite{tuckett2020fault}. They introduced a high-threshold decoder tailored specifically for a noise-tailored surface code. Their innovative approach, built upon minimum-weight perfect matching, was tailored to exploit symmetries in biased noise scenarios. While this approach achieved fault-tolerant thresholds exceeding 6\%, their research was primarily aimed at biased noise scenarios, without a broader examination of other noise conditions. In a later work, Tuckett et al. delved deeper into tailoring surface codes for scenarios dominated by highly biased noise, especially dephasing \cite{tuckett2019tailoring}. Their results revealed that with certain modifications, the surface code can achieve exceptionally high error-correction thresholds. Moreover, they provided evidence that the threshold error rate of the surface code closely aligns with the hashing bound across all biases. However, their focus remained on highly biased noise scenarios, not extending to a wider range of noise conditions.

The practical challenges of achieving low error rates in quantum computing were brought to the forefront by Google Quantum AI \cite{google2023suppressing}. Their work emphasized that while QECCs can indeed help in reducing error rates by encoding logical qubits within multiple physical qubits, this increase in physical qubits can also be a source of more errors. Nonetheless, using a superconducting qubit system and a distance-5 surface code logical qubit, they demonstrated the potential to overcome these additional error sources.

Krinner et al. focused on the realization of repeated quantum error correction by leveraging a distance-three surface code in a superconducting system with 17 physical qubits \cite{krinner2022realizing}. Their work aligned with ours in employing the STIM simulator, which further attested to its efficacy in producing results that closely mirror those from actual quantum computers.

In comparison to the aforementioned studies, our research offers a comprehensive perspective on QECCs. We delve deeper into the resilience of both repetition and surface codes, expanding the analysis to a diverse range of noise models and error types. Our findings provide richer insights into the performance and robustness of these codes, especially in the face of multiple types of errors, bridging some of the gaps observed in previous works.
\section{Examining Noise in Quantum Computing} \label{noise}


\subsection{Background}

In quantum computing, 'noise' refers to unwanted disturbances that can cause errors in qubits. There are mainly two error types: bit-flip (X-error) and phase-flip (Z-error) \cite{nielsen2001quantum}. Bit-flip errors invert a qubit's state, while phase-flip errors change the sign of the basis state $\ket{1}$. For a state $\ket{\psi} = \alpha\ket{0} + \beta\ket{1}$, a bit-flip results in $X\ket{\psi} = \alpha\ket{1} + \beta\ket{0}$, and a phase-flip gives $Z\ket{\psi} = \alpha\ket{0} - \beta\ket{1}$. These errors can combine, leading to complex disturbances.

\subsection{Varieties of Errors in Quantum Computing}

We offer a detailed exploration and modeling of errors in quantum computing, noting that errors only manifest when their probability is not zero.

\begin{enumerate}
\item \textit{Depolarizing error:} Arises from a qubit's interaction with the environment or thermal shifts. It randomizes the qubit state based on coherence time and quantum gate exposure duration. In our tests, we emulate this by applying the depolarizing probability during the start of a stabilizer measurement. A single-qubit depolarizing error picks and applies a random Pauli error (X, Y, Z). Maximum mixing happens at 75\% probability.
\item \textit{Gate error:} Emanates from quantum gate applications due to factors like miscalibration or environmental noise. We model gate errors in two segments: single and two-qubit Clifford operations. Each Clifford operation's execution considers the error probability. For two-qubit gate errors, a random pair of Pauli errors is applied, peaking the mixing at 93.75\% probability.
\item \textit{Readout error:} Occurs during qubit state measurements due to factors like detector inefficiencies. We model these errors as pre-measurement probabilistic operations, capturing the chance of X or Z flips.
\item \textit{Reset error:} Happens if a qubit, usually initialized to the $\ket{0}$ state, fails due to control issues or noise. We model these as post-reset probabilistic operations, applying potential X or Z flips based on the given probability.
\item \textit{Leakage error:} In systems with multiple levels, these errors happen when a qubit shifts out of its computational space, often due to faulty gate operations.
\item \textit{Crosstalk error:} Occur when a qubit operation unintentionally affects another qubit, typically from close physical proximity or unwanted interactions.
\end{enumerate}


\subsection{Modeling Noise for Quantum Computing}

\begin{figure}
    \centering
    \includegraphics[width=1\linewidth]{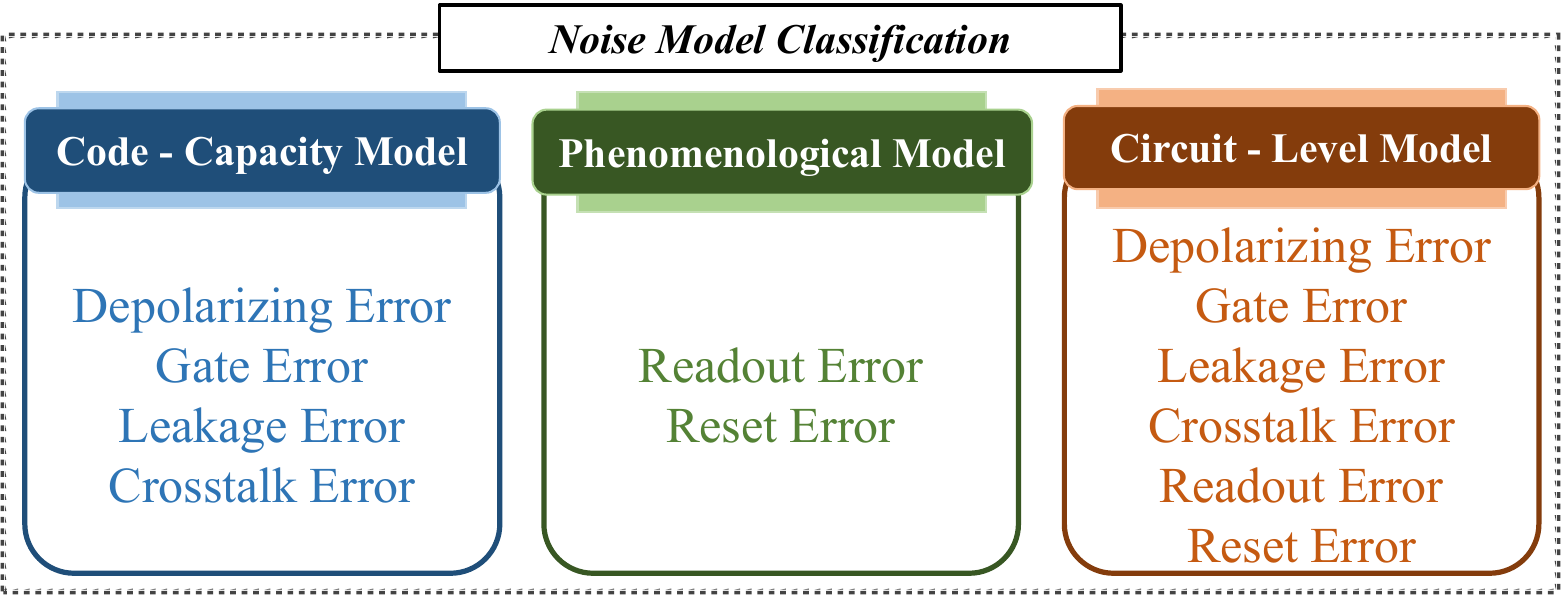}
    \caption{\textbf{Classification of noise models in our experiments.} The code-capacity model gauges error rate for data qubits, ignoring syndrome measurements. The phenomenological model estimates error thresholds with faulty syndrome measurements but disregards circuit-level error propagation. The circuit-level model comprehends errors in all operations, covering data and ancilla qubits.
    }
    \label{fig:noise_model}
    \vspace{-10pt}
\end{figure}

We've employed three noise models, inspired by the previously described noise types, to characterize QECCs. These models: code-capacity, phenomenological, and circuit-level, provide a comprehensive framework for various error scenarios in quantum computing \cite{li20192d, zhang2023concatenation} (Fig. \ref{fig:noise_model}).

\begin{enumerate}
\item \textit{Code-capacity model:} Assumes flawless measurements. It integrates a depolarizing probability at each round's start and a random gate error after every Clifford operation. This model, capturing both pre- and post-operation phases, gauges the full resilience of the error-correcting code.
\item \textit{Phenomenological noise model:} Assumes perfection in all but measurements. This model includes the probability of a qubit undergoing an X or Z flip pre-measurement and post-reset.
\item \textit{Circuit-level noise model:} Acknowledges potential imperfections throughout the quantum system, covering quantum gates, initialization, and measurements. It's the most comprehensive model, accounting for various errors: depolarizing probability at each round's inception, gate errors post Clifford operations, and qubit flips both pre-measurement and post-reset. This thorough approach provides a granular insight into each quantum operation's potential error contribution.
\end{enumerate}

\subsection{Relevance of the Noise Models}

A QECC comprises two main components: the actual circuit involving logical qubits \& gates, and syndrome measurements featuring ancilla qubits \& stabilizer measurements.

The code-capacity model evaluates the maximum error rate a code can correct, focusing exclusively on the actual circuit and its data qubits, disregarding the syndrome measurement part. This model also evaluates the error propagation in the data qubits over time, covering the impact of cumulative errors on the overall computation. Contrarily, the phenomenological model estimates the accuracy threshold considering faulty syndrome measurements. It scrutinizes the ancilla qubits and the repeated stabilizer measurements, overlooking the detailed circuit-level error propagation. Lastly, the circuit-level model encapsulates errors across all elementary operations in the entire circuit, including data and ancilla qubits. It simulates a realistic setting where all operations are noisy and provides an estimation of the accuracy threshold considering error propagation within the circuit \cite{landahl2011fault}.

Using these noise models, we can compare various QECCs under equivalent conditions, allowing fair assessments of their intrinsic error tolerance.

\subsection{Experimental Framework} 

We examined two QECCs, the rotated and unrotated surface codes, under the noise types and models previously outlined. Table \ref{tab:all_qecc_details} showcases QECC features, including distance, qubit count, and gates used. As distance and rounds grow, so do qubit and gate overheads. Using the equation $rounds = distance \times 3$, we visualize computation as a 3D structure: one $d \times d$ lattice per round, for a total height of $3d$. All displayed logical error rates are per round, decoded using the MWPM approach \cite{kolmogorov2009blossom, higgott2022pymatching}.

In this study, we conduct our experiments using STIM \cite{gidney2021stim}, a high-speed simulator for quantum stabilizer circuits. STIM leverages a stabilizer tableau representation, akin to the CHP simulator \cite{aaronson2004improved}, but also exhibits significant enhancements for performance and accuracy. As corroborated by the paper \cite{krinner2022realizing}, STIM offers an unparalleled level of fidelity in emulating the operation of QECCs on actual quantum hardware. We configured STIM simulations to evaluate repetition and surface codes under different distances, rounds, noise types, noise models, and physical error rates. The noise models included depolarizing, gate, readout, and reset errors with controlled probabilities to capture diverse error scenarios. Logical error rates were extracted from the simulator outputs across the sampled physical error rates. These values were used to generate plots relating logical and physical error rates for each code and noise configuration. By analyzing the logical error rate curves, we identified accuracy thresholds where the QECCs begin failing. Comparing thresholds and resource overheads allowed for assessing the relative performance and limitations of the codes. This simulation-based workflow enabled a comprehensive evaluation of the noise resilience and optimal implementation of QEC codes to match target logical error rates and hardware constraints.

Given the present constraints and limitations in the number of qubits of real quantum hardware (freely accessible ones), the execution of QECCs directly on these platforms is not feasible. Hence, STIM proves to be the best tool for our research.

\begin{table}[]
\caption{Characteristics of QECCs in Our Experiments}
\centering
\begin{tabular}{c|c|c|c}
\hline
\multicolumn{1}{c|}{\textbf{QECC Type}} & \textbf{Dist.}& \textbf{\#Qb} & \textbf{\#Gates} \\ \hline \hline
\multirow{4}{*}{Unrotated Surface}  & 3                 & 25                 & 104               \\  
                                         & 5                 & 81                 & 328               \\ 
                                         & 7                 & 169                & 792               \\  
                                         & 9                 & 289                & 1376              \\ \hline 
\multirow{4}{*}{Rotated Surface}    & 3                 & 17                 & 64                \\  
                                         & 5                 & 64                 & 208               \\  
                                         & 7                 & 118                & 432               \\  
                                         & 9                 & 170                & 736               \\ \hline \hline
\end{tabular}
\label{tab:all_qecc_details}
\vspace{-10pt}
\end{table}

\section{Analyzing Surface Codes Under Noise} \label{all_codes}

In this section, we examine the resilience and efficiency of rotated and unrotated surface codes under various types of noise, including depolarizing, gate, readout, and reset errors. We also consider different noise models. 

\subsection{Surface Codes under Various Noise Types}

\begin{figure*}
    \centering
    \includegraphics[width=1\linewidth]{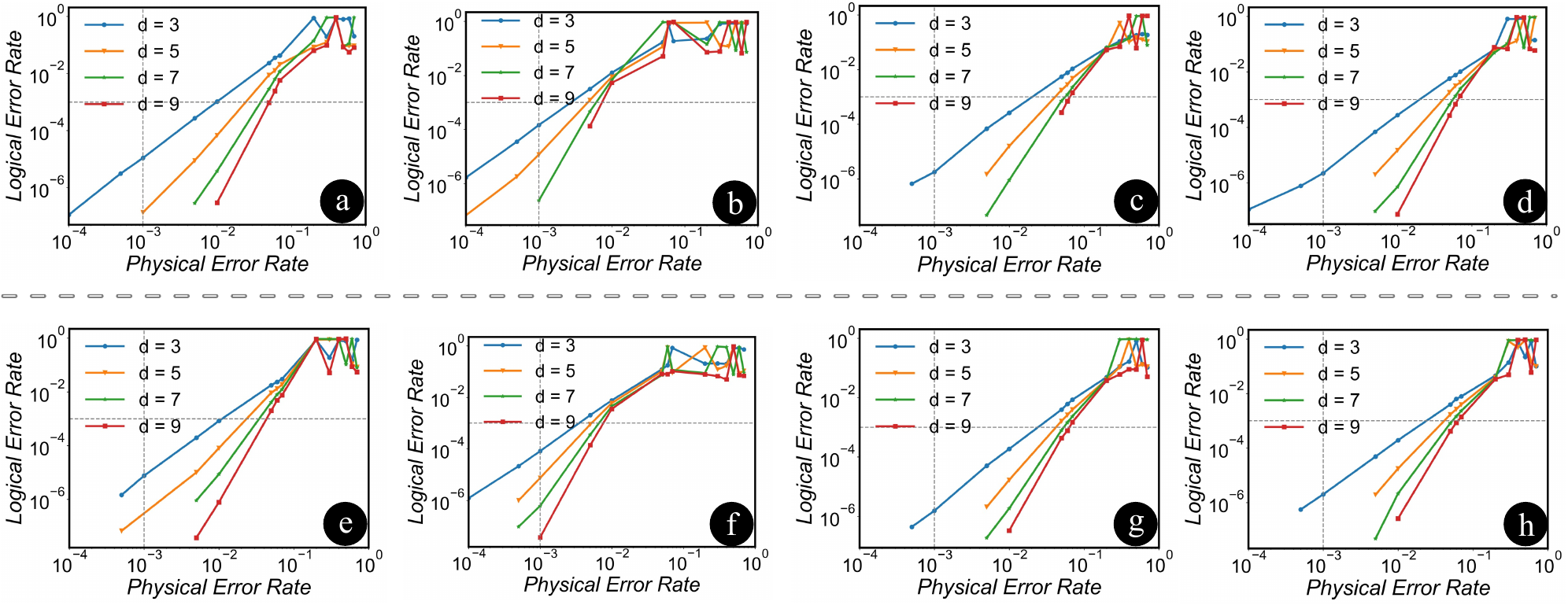}
    \caption{\textbf{Logical error rate vs. physical error rate in unrotated and rotated surface codes with varying distances for various error types}. The black dotted line marks the ~$10^{-3}$ error rate of modern processors. Errors include: \raisebox{.5pt}{\textcircled{\raisebox{-.4pt} {\textbf{a}}}} Depolarizing; \raisebox{.5pt}{\textcircled{\raisebox{-.9pt} {\textbf{b}}}} Gate; \raisebox{.5pt}{\textcircled{\raisebox{-.2pt} {\textbf{c}}}} Reset; \raisebox{.5pt}{\textcircled{\raisebox{-.9pt} {\textbf{d}}}} Readout for unrotated surface codes; with \raisebox{.5pt}{\textcircled{\raisebox{-.4pt} {\textbf{e}}}} - \raisebox{.5pt}{\textcircled{\raisebox{-.9pt} {\textbf{h}}}} representing similar experiments with rotated surface codes. 
    }
    \label{fig:surface_all_noise}
    \vspace{-10pt}
\end{figure*}

Initially, we delve into how the logical error rate diminishes in response to an increase in unrotated surface code distances across a spectrum of physical error rates if the hardware is dominated by either depolarizing, gate, readout, or reset errors. For depolarizing noise (Fig. \ref{fig:surface_all_noise} \raisebox{.5pt}{\textcircled{\raisebox{-.4pt} {\textbf{a}}}}), we note that increasing the distances of the unrotated surface code can effectively reduce the logical error rate, indicating the improvement in system's error correction capabilities. Nevertheless, this improvement is only significant up to a certain point, beyond which code distance increment fails to yield substantial benefits. This point is known as the threshold of the error-correcting code at which the physical error rate is so high that the QECC loses its efficacy, and the system becomes inherently unreliable. 
When rotated surface codes are being examined under depolarizing noise as shown in Fig. \ref{fig:surface_all_noise} \raisebox{.5pt}{\textcircled{\raisebox{-.4pt} {\textbf{e}}}} respectively, we note similar trends such as the logical error rate reduces for higher code distances up to the threshold level.

Likewise, if we prioritize the impact of gate errors as the dominant factor, the logical error rate shows similar trends when analyzed using different distances of QECCs. However, in this case, the error correction threshold is lower than the depolarizing errors. This observation holds true for unrotated (Fig. \ref{fig:surface_all_noise} \raisebox{.5pt}{\textcircled{\raisebox{-.9pt} {\textbf{b}}}}) and rotated (Fig. \ref{fig:surface_all_noise} \raisebox{.5pt}{\textcircled{\raisebox{-.9pt} {\textbf{h}}}}) surface codes. This can be explained by the fact that gate errors occur during the execution of quantum gate operations. These errors can lead to a substantial accumulation of errors over time, particularly for larger and more complex quantum circuits. In contrast, depolarizing errors are modeled as a probability that occurs before each round of computations. As such, they do not directly interfere with the gate operations themselves, and their propagation effect is thus generally lower than gate errors.

The trend of logical error rate reduction with increasing unrotated surface code distance applies to readout and reset errors as well, as depicted in Figs. \ref{fig:surface_all_noise} \raisebox{.5pt}{\textcircled{\raisebox{-.2pt} {\textbf{c}}}} and \raisebox{.5pt}{\textcircled{\raisebox{-.9pt} {\textbf{d}}}}. However, the key distinction lies in the higher thresholds of readout and reset errors as compared to both depolarizing and gate errors. This can be attributed to the lack of propagation effects associated with readout and reset errors, which are primarily linked with the measurement and initialization processes in quantum computing. To be more specific, readout and reset errors, despite producing incorrect results, do not accumulate. 

All of these characteristics hold true for rotated surface codes (Fig. \ref{fig:surface_all_noise} \raisebox{.5pt}{\textcircled{{\textbf{g}}}} \& \raisebox{.5pt}{\textcircled{\raisebox{-.9pt} {\textbf{h}}}}).

For further clarification, we have marked the general error rate of a state-of-the-art quantum processor, which is approximately $10^{-3}$, in all the plots in Fig. \ref{fig:surface_all_noise} using a horizontal and a vertical black dotted line. 
These lines indicate that both the physical and logical error rates would be situated at or below this level without the application of quantum error correction. By applying QECC, we can significantly decrease the logical error rate below this level and as the distance of the codes is increased, the system will become more fault-tolerant.

\begin{figure}
    \centering
    \includegraphics[width=1\linewidth]{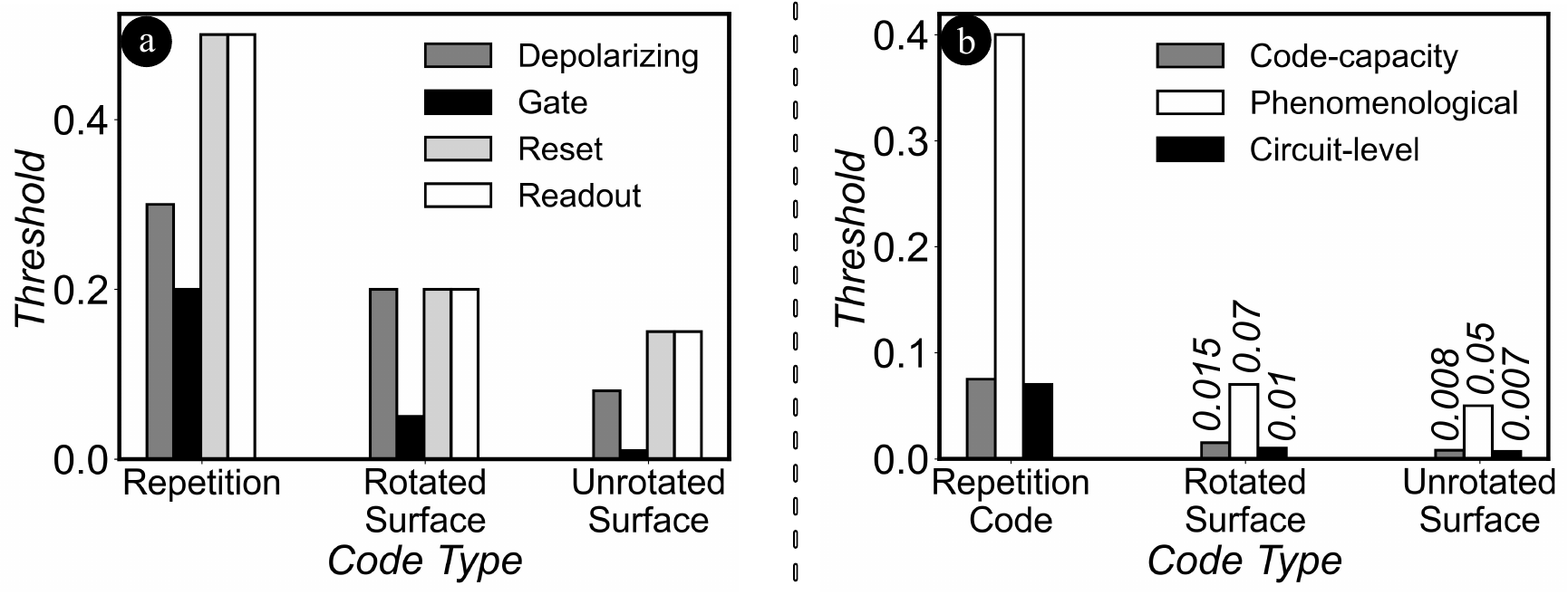}
    \caption{\textbf{Comparison between threshold values of repetition and surface codes} for \raisebox{.5pt}{\textcircled{\raisebox{-.4pt} {\textbf{a}}}} different error types; \raisebox{.5pt}{\textcircled{\raisebox{-.9pt} {\textbf{b}}}} various noise models. While the threshold of repetition codes is markedly higher than that of surface codes, this is largely due to its inability to detect or correct Z errors, thus ignoring half of potential errors. 
    The threshold of the rotated surface code is better than the unrotated surface code. 
    } 
    \label{fig:why_rep_bar_graph}
    \vspace{-10pt}
\end{figure}

Fig. \ref{fig:why_rep_bar_graph} shows rotated surface codes outperforming unrotated ones due to their reduced complexity and fewer accumulated errors, as detailed in Table \ref{tab:all_qecc_details}. This is also due to fewer accumulated errors from a lower gate count. Hence, from the subsequent subsection onward, we confine our experiments to rotated surface codes exclusively. While repetition codes show higher thresholds, they can't detect or correct Z errors. This limitation makes them suitable only for specific cases with dominant bit-flip errors, whereas surface codes effectively address both X and Z errors.

\subsection{Surface Codes Under Various Noise Models}

\begin{figure}
    \centering
    \includegraphics[width=1\linewidth]{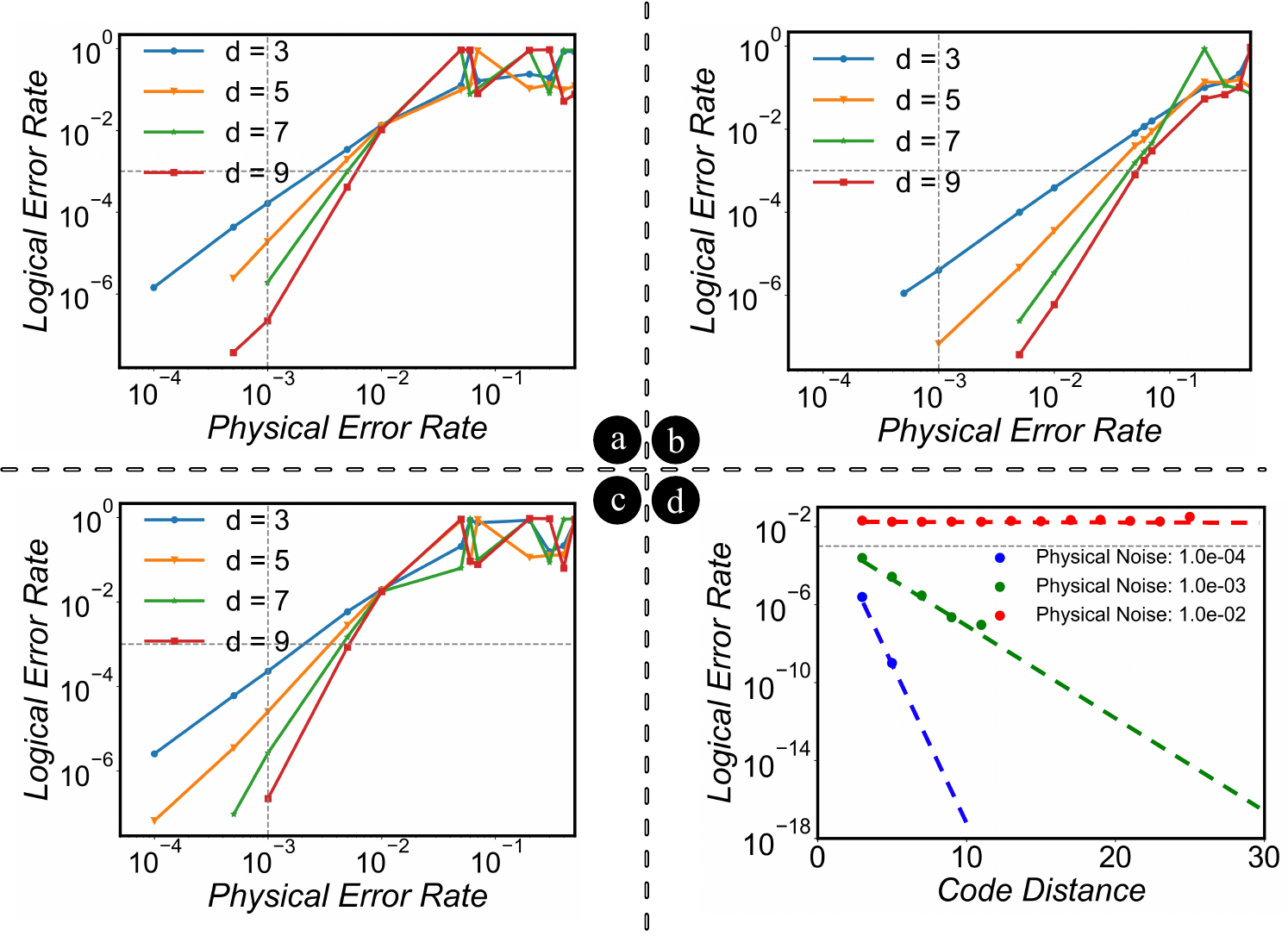}
    \caption{\textbf{Logical error rate vs. physical error rate in rotated surface codes} for \raisebox{.5pt}{\textcircled{\raisebox{-.4pt} {\textbf{a}}}} Code-capacity noise, \raisebox{.5pt}{\textcircled{\raisebox{-.9pt} {\textbf{b}}}} Phenomenological noise, and \raisebox{.5pt}{\textcircled{\raisebox{-.2pt} {\textbf{c}}}} Circuit-level noise models. These plots reveal the relationship between logical and physical error rates under different error models, with the surface code's distance as a variable.  
    \raisebox{.5pt}{\textcircled{\raisebox{-.9pt} {\textbf{d}}}}: \textbf{Logical error rate vs. code distance in rotated surface code} depicting the superior performance of QECCs at lower physical noise levels. A threshold physical noise of $10^{-2}$ is observed, beyond which, despite increasing the distance, QECC performance ceases to improve due to excessive physical noise. The black dotted line represents the typical error rate of modern processors.
    } 
    \label{fig:surface_all_model}
    \vspace{-10pt}
\end{figure}

Next, we analyze the performance of surface codes under three distinct noise models as shown in Fig. \ref{fig:surface_all_model}, respectively. 
Beginning with the code-capacity model (Fig. \ref{fig:surface_all_model} \raisebox{.5pt}{\textcircled{\raisebox{-.4pt} {\textbf{a}}}}), 
it can be discerned that the threshold for this model is similar to that of gate errors. This similarity arises since the threshold of this model is determined by the lower of the two error types, which in this case, is the gate errors. Moving on to the phenomenological model (Fig. \ref{fig:surface_all_model} \raisebox{.5pt}{\textcircled{\raisebox{-.9pt} {\textbf{b}}}}), which is only about the syndrome measurements in ancilla qubits, exhibits high similarity with the plots of readout and reset errors. The threshold condition for this model also aligns closely with these two measurement-type errors. 
Finally, the circuit-level (Fig. \ref{fig:surface_all_model} \raisebox{.5pt}{\textcircled{\raisebox{-.2pt} {\textbf{c}}}}) noise model encompasses all types of errors, across the QECC including both data and ancilla qubits, making it the most representative of real-world scenarios. Accordingly, the threshold for these combined errors is the lowest among all the thresholds we have examined so far.  

We also mapped the logical error rate against code distance while varying physical noise levels in rotated surface codes for the circuit-level noise model (which is the most realistic scenario) as depicted in Fig. \ref{fig:surface_all_model} \raisebox{.5pt}{\textcircled{\raisebox{-.9pt} {\textbf{d}}}}. We utilized a linear regression model to fit a line to the sampled data, establishing a relationship between the code distance and the logical error rate under various physical noise levels. This enabled us to predict the code distance necessary to achieve a specified logical error rate in the code-capacity model. The plots indicate that a reduction in physical noise directly correlates to a steeper slope. 
In other words, when the quantum system is inherently less noisy, enhancing the code distance has a pronounced impact in diminishing the logical errors. Hence, it is essential to consider the native noise level of the quantum system when optimizing code distance for error correction. However, we identified a noise level threshold, beyond which the performance improvement of QECCs plateaus, regardless of distance. It is noteworthy that the lowest error threshold that surface code can achieve is around $10^{-2}$, a figure that is 10 times higher than what is achievable by state-of-the-art quantum processors without error correction. This underscores the pivotal role of QECCs in the advancement of quantum computing.

\section{Strategies for Optimal \& Scalable Code Selection} \label{advancing_eff}

This section will synthesize the key insights from the noise characterization study to outline effective strategies for selecting optimal quantum error correction codes and implementing them in a tailored manner to match quantum hardware constraints. It will provide guiding principles to help determine the most suitable code, distance, rounds, and qubit resources based on the predominant error types, target error rates, and hardware limitations. The discussion will focus on balancing performance, overhead, and practical viability when designing QECC architectures. It will propose techniques to avoid both underestimating and overestimating code parameters based on projected application requirements and hardware capabilities. 

\subsection{Performance Analysis of Surface Codes with Varying Rounds}

\begin{figure}
    \centering
    \includegraphics[width=1\linewidth]{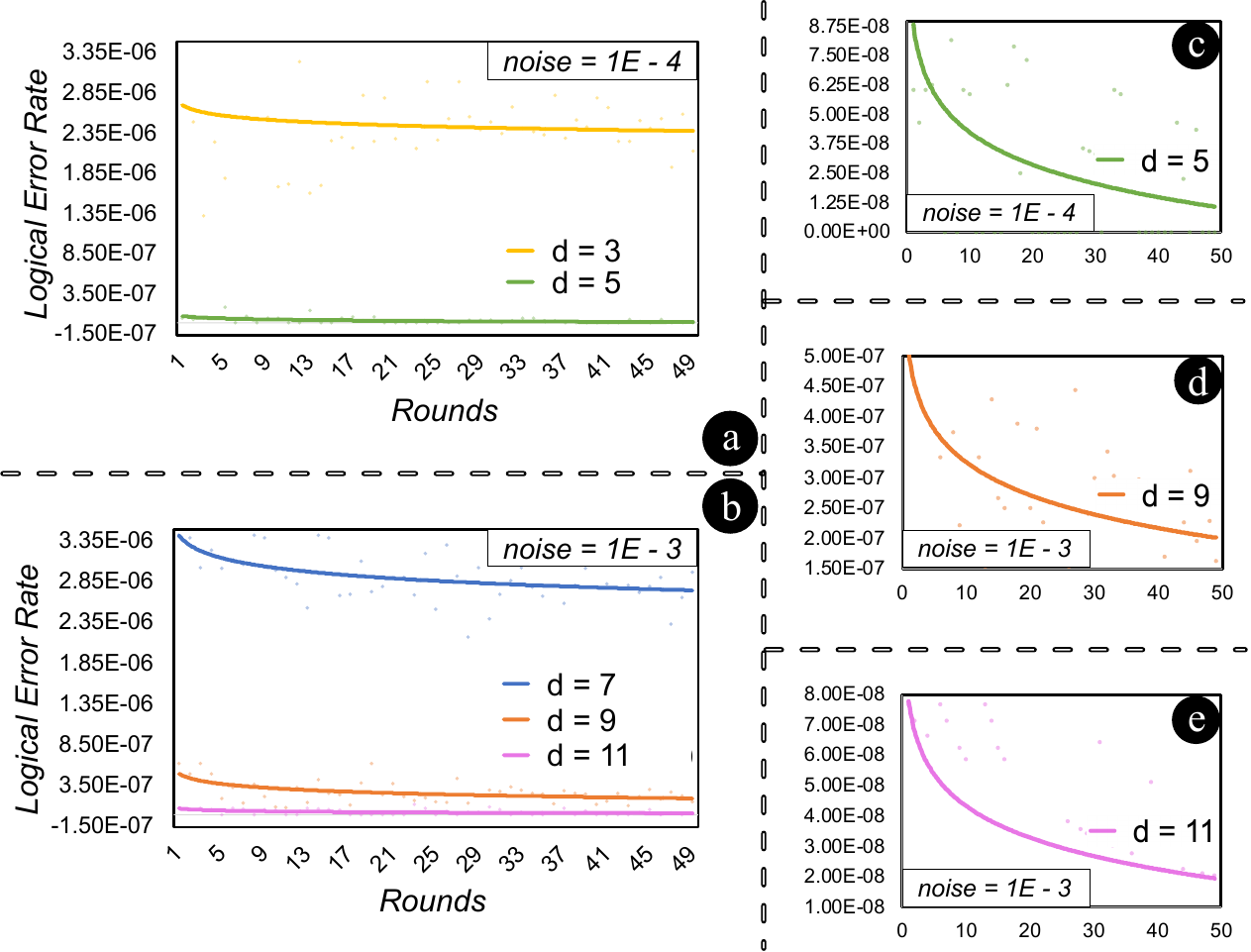}
    \caption{\textbf{Logical error rate vs. number of rounds for rotated surface codes} of \raisebox{.5pt}{\textcircled{\raisebox{-.4pt} {\textbf{a}}}} distance $3$ and $5$ under physical noise level of $10^{-4}$. \raisebox{.5pt}{\textcircled{\raisebox{-.9pt} {\textbf{b}}}} distance $7$, $9$ and $11$ under physical noise level of $10^{-3}$.  Subfigures \raisebox{.5pt}{\textcircled{\raisebox{-.2pt} {\textbf{c}}}}, \raisebox{.5pt}{\textcircled{\raisebox{-.9pt} {\textbf{d}}}}, and \raisebox{.5pt}{\textcircled{\raisebox{-.2pt} {\textbf{e}}}}, show magnified views for distances $5$, $9$, and $11$.} 
    \label{fig:vary_rounds}
    \vspace{-10pt}
\end{figure}

To begin with, we study the performance of surface code in relation to rounds (which denotes the number of times an error correction code is executed and stabilizers are measured). Theoretically, the code becomes more efficient with increasing number of rounds. In Fig. \ref{fig:vary_rounds}, we apply two sets of physical noise levels: $10^{-4}$ and $10^{-3}$ and execute distance 3 and 5 rotated surface codes for the first set, and distance 7, 8, and 9 for the second set. We fit a logarithmic curve to the collected data to study its pattern and behavior. The choice to vary the physical noise levels between two different sets of distances is predicated upon the results from Fig. \ref{fig:surface_all_noise} and Fig. \ref{fig:surface_all_model}. A physical noise of $10^{-4}$ is relatively low for a surface code $distance \geq 7$, while a physical noise of $10^{-3}$ would prove excessive for $distance \leq 5$.

Fig. \ref{fig:vary_rounds} \raisebox{.5pt}{\textcircled{\raisebox{-.4pt} {\textbf{a}}}} illustrates the behavior of the logical error rate as we vary the rounds of a distance 3 and 5 rotated surface code. As anticipated, the logical error rate decreases with each additional round. This is because the measure used in our plots is the logical error rate per round, meaning that as rounds increase, the effectiveness of error detection and correction also increases, thereby reducing the error rate per round. Likewise, in \raisebox{.5pt}{\textcircled{\raisebox{-.9pt} {\textbf{b}}}}, we demonstrate that the logical error rate reduces with an increment in rounds for distances 7, 9, and 11 rotated surface codes, even though the noise level is higher than the previous set. Fig. \ref{fig:vary_rounds} \raisebox{.5pt}{\textcircled{\raisebox{-.2pt} {\textbf{c}}}}, \raisebox{.5pt}{\textcircled{\raisebox{-.9pt} {\textbf{d}}}}, and \raisebox{.5pt}{\textcircled{\raisebox{-.2pt} {\textbf{e}}}} display magnified plots for distances 5, 9, and 11 to provide a clear visualization of the line curvatures, reinforcing our understanding that higher distances of a surface code with more rounds perform better in a given scenario.
However, this improvement comes with an overhead in terms of the number of qubits and gate count as shown in Table \ref{tab:all_qecc_details}.

\subsection{Resource Requirement for Fault Tolerance}

Our findings prompt pivotal questions: Is it always essential to operate surface codes at their maximum distance and rounds? Can we design adaptable surface codes with dynamic parameters for various noise types and models? 

In reality, the achievement of true fault tolerance involves maintaining a low, albeit non-zero, error rate. For a quantum computer dominated by specific types of errors or noise models, understanding the system's targeted logical error rate and noise characteristics allows us to optimize surface code parameters. This will reduce unnecessary qubit and gate overhead. Different quantum hardware may experience varying types of errors. Consequently, it is crucial to customize QECC parameters, i.e. distance and number of qubits, to optimally benefit each unique system. The Table \ref{tab:fault_tolerant} provides the minimum distance and qubit requirements for achieving fault tolerance (logical error rate $< 10^{-9}$) at a physical noise level of $10^{-3}$ using rotated surface codes.

\begin{table}[]
\centering
\caption{Minimum Resource Requirement for Fault Tolerance at Physical Noise $= 10^{-3}$ in Surface Codes}
\begin{tabular}{l|c|c}
\hline
\multicolumn{1}{c|}{\textbf{Error or Noise Model}} & \textbf{Min.Dist.} & \textbf{\#Qb} \\ \hline \hline
Depolarizing error                                  & 7                         & 118                \\ 
Gate error                                          & 9                         & 170                \\ 
Reset error                                         & 5                         & 64                 \\ 
Readout error                                       & 5                         & 64                 \\ \hline
Code-capacity model                                 & 7                         & 118                \\ 
Phenomenological model                              & 5                         & 64                 \\ 
Circuit-level model                                 & 9                         & 170                \\ \hline \hline
\end{tabular}
\label{tab:fault_tolerant}
\vspace{-15pt}
\end{table}

As shown in Table \ref{tab:fault_tolerant}, the resource demands for implementing surface codes differ based on the dominant error type in a quantum hardware system. Among the errors, reset \& readout errors are the least resource-intensive, followed by depolarizing errors, and lastly, gate errors, which require the most resources. 
This ranking is consistent across various types of quantum codes, including repetition codes, unrotated surface codes, and rotated surface codes.

Similar to the error types, when a specific noise model predominantly characterizes quantum hardware, there is a corresponding ranking in terms of resource requirements for surface code implementation. The phenomenological model requires the fewest resources, followed by the code-capacity model, and finally, the circuit-level model, which is the most resource-demanding. This order is universal and applies to repetition, unrotated, and rotated surface codes alike.


\subsection{Performance Analysis of Surface Codes with Code Distance}

\begin{figure}
    \centering
    \includegraphics[width=1\linewidth]{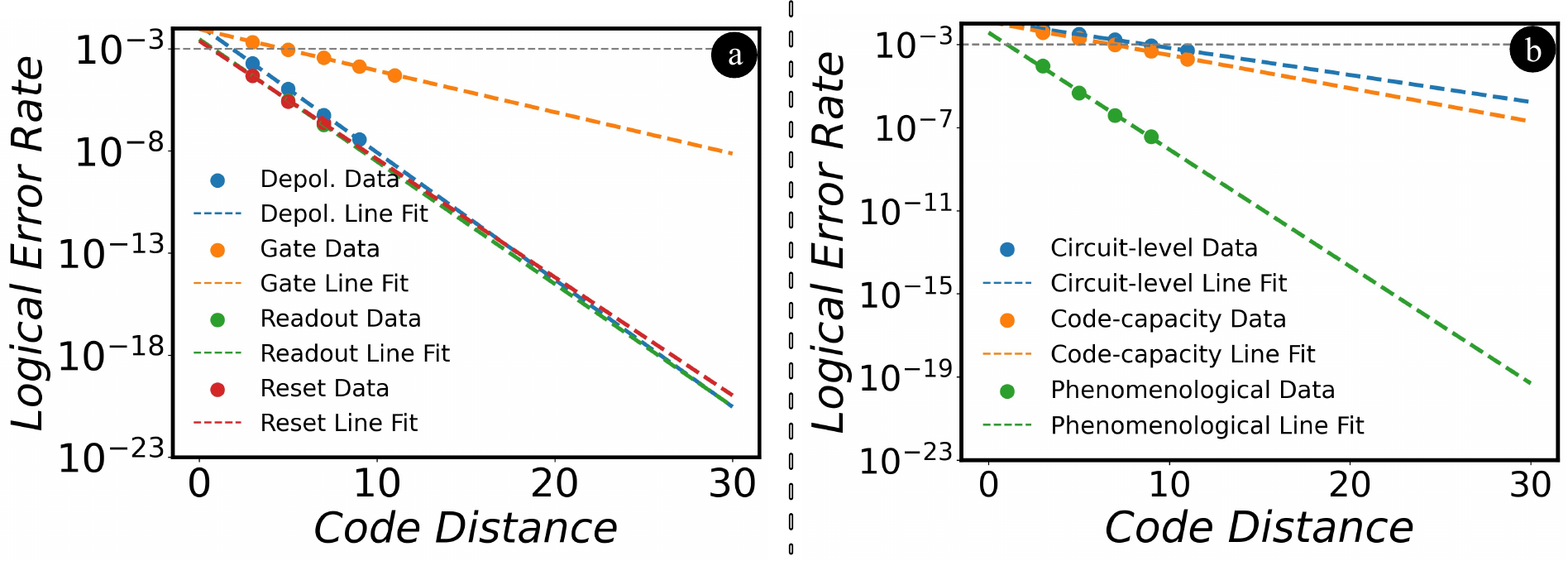}
    \caption{\textbf{Logical error rate vs. code distance for rotated surface codes} at a physical noise level of $0.5 \times 10^{-2}$ for \raisebox{.5pt}{\textcircled{\raisebox{-.4pt} {\textbf{a}}}} various error types and \raisebox{.5pt}{\textcircled{\raisebox{-.9pt} {\textbf{b}}}} noise models. The experiments are projected to a distance $30$ with $rounds = distance \times 3$. The results confirm that optimal surface code is not always needed. By understanding the target logical error rate for certain noise types or models, we can identify appropriate surface code parameters, reducing unnecessary gate and qubit overhead.
    } 
    \label{fig:surface_fitting}
    \vspace{-10pt}
\end{figure}

We conducted an experiment to determine the amount of logical errors produced when certain parameters of a surface code are applied at a fixed physical noise level. We used a linear regression model to plot a line fit of code distance against the log of the logical error rate. By plotting the collected points alongside the line fit, we projected the distance required to achieve the desired logical error rate. This experiment was executed in two sets: one accounting for different errors, and the other considering various noise models. We maintained the physical noise at $0.5 \times 10^{-2}$, a value that exceeds the noise level of current state-of-the-art technology. For these experiments, we set $rounds = distance \times 3$ and varied the distance from $3$ to $9$ to extrapolate the line fit till distance $30$. The results of these experiments are exhibited in Fig. \ref{fig:surface_fitting} \raisebox{.5pt}{\textcircled{\raisebox{-.4pt} {\textbf{a}}}} and \raisebox{.5pt}{\textcircled{\raisebox{-.9pt} {\textbf{b}}}}. In our subsequent experiment, as depicted in Figure \ref{fig:vary_qubit_fit}, we also assessed the number of qubits required to attain a specified logical error rate for rotated surface codes. In this assessment, we extended our projections to consider systems utilizing up to $1000$ qubits. This step was crucial as it allowed us to observe the required resource for the targeted logical error rate, which varies based on different error and noise models. The following conclusions can be drawn:

Even with the most realistic Circuit-level noise model, an elevated distance yields a notably low logical error rate at such high physical error levels. As exemplified in Fig. \ref{fig:surface_fitting} \raisebox{.5pt}{\textcircled{\raisebox{-.4pt} {\textbf{a}}}}, if we are dealing with a quantum hardware system predominantly affected by depolarizing errors and we aim to achieve a logical error rate of $10^{-10}$, a surface code of approximately distance $12$ with $48$ rounds $(12 \times 3)$ will be necessary. Conversely, if the system is more heavily affected by gate errors, a surface code exceeding a distance of $30$ (and rounds 90) would be required to achieve the same logical error rate.

\subsection{Assessing Surface Codes Relative to Current State-of-the-Art Quantum Processors}

\begin{figure}
    \centering
    \includegraphics[width=1\linewidth]{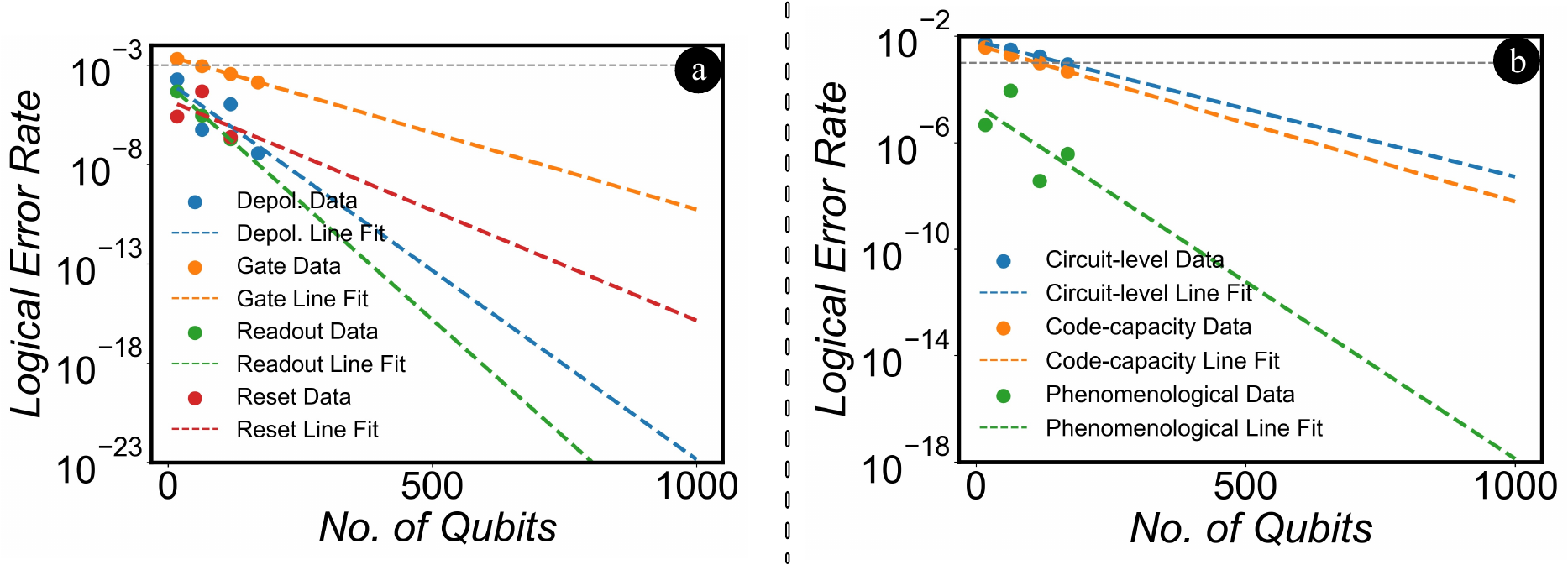}
    \caption{\textbf{Logical error rate vs. number of qubits for rotated surface codes} at a physical noise level of $0.5 \times 10^{-2}$ for \raisebox{.5pt}{\textcircled{\raisebox{-.4pt} {\textbf{a}}}} various error types and \raisebox{.5pt}{\textcircled{\raisebox{-.9pt} {\textbf{b}}}} noise models. The experiments are projected to $1000$ qubits. The findings reveal that the proper determination of surface code parameters can effectively balance the logical error rate and qubit overhead.
    } 
    \label{fig:vary_qubit_fit}
    \vspace{-10pt}
\end{figure}

While minimizing the logical error rate, we must also consider qubit overhead, which varies with error types and noise models. As exemplified in Fig. \ref{fig:vary_qubit_fit} \raisebox{.5pt}{\textcircled{\raisebox{-.4pt} {\textbf{a}}}}, if we are dealing with a quantum hardware system predominantly affected by depolarizing errors, along with a physical noise of $0.5 \times 10^{-2}$ (higher than the current state-of-the-art error rates) and a target logical error rate of $10^{-23}$, we will need approximately $1000$ qubits. Reducing the logical error rate further may be infeasible due to qubit cost and scarcity. Thus, identifying the right surface code parameters can help balance between the logical error rate and qubit overhead. Table \ref{tab:improvement} illustrates the interplay between code parameters (distance \& rounds), overhead (qubit \& gate counts), and system improvement for a surface code in the most realistic scenario: a circuit-level noise model with a physical error rate of $10^{-3}$. It is apparent that the system's performance significantly improves, up to $10^{12}$ times, with an increase in resources, reaching this point with a distance $25$ rotated surface code employing $580$ qubits and $2040$ gates. Continuous resource augmentation can indeed boost error rate improvement in quantum error correction codes. However, constraints, whether they be technological or logistical, impose a ceiling on resource availability.

\begin{table}[t]
\centering
\caption{System Improvement by Surface code at a Physical error rate of $10^{-3}$ under a Circuit-level noise model}
\begin{tabular}{cc|cc|c}
\hline
\multicolumn{2}{c|}{\textbf{Parameters}}                                      & \multicolumn{2}{c|}{\textbf{Overhead}}                                           & \multirow{2}{*}{\textbf{Improvement}} \\ \cline{1-4} 
\multicolumn{1}{c|}{\textbf{Dist.}} & \multicolumn{1}{c|}{\textbf{R}} & \multicolumn{1}{c|}{\textbf{\#Qb}} & \multicolumn{1}{c|}{\textbf{\#G}} &                                       \\ \hline \hline
3                 & 9               & 17                 & 64                & $\sim 10$ \textit{times}        \\ 
5                 & 15              & 64                 & 208               & $\sim 10^{2}$ \textit{times}    \\ 
7                 & 21              & 118                & 432               & $\sim 10^{3}$ \textit{times}    \\ 
11                & 33              & 170                & 736               & $\sim 10^{4}$ \textit{times}    \\ 
13                & 39              & 271                & 920               & $\sim 10^{5}$ \textit{times}    \\ 
15                & 45              & 323                & 1144              & $\sim 10^{7}$ \textit{times}    \\ 
17                & 51              & 374                & 1368              & $\sim 10^{8}$ \textit{times}    \\ 
21                & 63              & 477                & 1592              & $\sim 10^{9}$ \textit{times}    \\ 
23                & 69              & 528                & 1816              & $\sim 10^{10}$ \textit{times}   \\ 
25                & 75              & 580                & 2040              & $\sim 10^{12}$ \textit{times}   \\ \hline \hline
\end{tabular}
\label{tab:improvement}
\vspace{-10pt}
\end{table}


While it is natural to strive for the most efficient surface code utilization in every situation, this might not always be necessary or even feasible. It is crucial to consider the specific requirements and constraints of each scenario. By understanding the target logical error rate and the available resources in terms of qubit overhead, we can tailor our approach to the specific type of noise or noise model present. This way, we can determine the parameters of a rotated surface code that are best suited for the situation at hand.

Our analysis of QEC codes highlights that while surface codes effectively safeguard quantum data, they come with significant overhead. For instance, a distance-5 rotated surface code needs about 60 qubits, a challenge given the limited qubits in today's quantum computers. Moreover, the current hardware struggles with the precision needed for two-qubit gates. Despite these challenges, the promising error tolerance thresholds suggest surface codes are attainable. The focus should be on increasing both the number and quality of qubits. Enhancing qubit stability and gate fidelity will push error rates below critical thresholds. Optimizing decoding methods and crafting efficient codes will further bridge the gap between existing hardware and QEC demands, paving the way for reliable quantum computing.
\section{Limitations and Future Work} \label{limitaions}

This study offers a robust analysis of QEC codes under prevalent noise models, yet there are opportunities for further research due to some limitations. For example, we did not incorporate amplitude and phase-damping noise models, which arise from energy dissipation and dephasing — common phenomena in quantum systems. Modeling these necessitates moving beyond Pauli errors to density matrices and Kraus operator sums. The STIM simulator we employed is tailored for Pauli channels, omitting non-Pauli noise processes. While depolarizing noise offers some insight, directly incorporating amplitude and phase damping could alter our findings, especially regarding surface codes that counteract phase errors. Future work could integrate STIM with simulators supporting damping noise models, building on established quantum programming structures. Preliminary studies hint that our core findings might persist with these realistic noise models, but thorough assessments are crucial.

Our study also sidestepped leakage and crosstalk noise models, which can move qubits beyond the computational basis or produce correlated multi-qubit errors. Understanding these complex noises is essential as quantum hardware advances. They require more intricate quantum error models, transcending the simplistic single-qubit Pauli error assumption. A potential approach could be tensor network representations that encapsulate multi-qubit correlations. Merging such noise models with STIM might shed light on leakage and crosstalk effects. While analytical models offer a starting point, simulations that capture these errors' nuances might be indispensable.
Future research avenues include exploring a broader spectrum of codes, refining decoders, investigating imperfect syndrome measurements, and delving into code concatenation. 

\section{Conclusion} \label{conclusion}

This research offers pivotal insights for the tailored design and execution of QEC codes in line with the distinct noise challenges of modern quantum hardware. We have established that surface codes excel in quantum error protection, proficiently managing both common bit-flip and phase-flip errors, a contrast to basic repetition codes which only cater to bit-flips. Among the surface code variations, rotated surface codes emerge superior due to their higher error thresholds, simplicity, and efficient qubit usage. Such findings guide researchers in choosing the most apt QECC structure, harmonizing with desired reliability and hardware constraints. Our extensive exploration of noise models, ranging from the stringent code-capacity to the pragmatic circuit-level models, offers a detailed portrayal of QECC durability. 
The study also underscores the value of fine-tuning QECC parameters, such as distance, to strike a balance between desired reliability and qubit economy. In essence, this research is a significant stride towards achieving scalable, fault-tolerant quantum computing, laying down the foundation for QECC designs attuned to specific quantum noise scenarios. 
\section*{Acknowledgements}
The work is supported in parts by the National Science Foundation (NSF) (CNS-1722557, CCF-1718474, OIA-2040667, DGE-1723687, and DGE-1821766).

\bibliographystyle{plainnat} 

\bibliography{refs}

\end{document}